\documentclass[journal=jacsat,manuscript=article]{achemso}

\usepackage[version=3]{mhchem}

\usepackage{amsmath}
\usepackage{xcolor}
\usepackage{ulem}
\usepackage{comment}

\author{Zhengxin Li}
\affiliation{Department of Chemical and Materials Engineering, University of Alberta, Alberta T6G 1H9, Canada}
\author{Akihito Kiyama}
\affiliation{Department of Mechanical and Aerospace Engineering, Utah State University, Logan, UT 84322, USA}
\author{Hongbo Zeng}
\affiliation{Department of Chemical and Materials Engineering, University of Alberta, Alberta T6G 1H9, Canada}
\email{hongbo.zeng@ualberta.ca}
\author{Xuehua Zhang}
\email{xuehua.zhang@ualberta.ca}
\affiliation{Department of Chemical and Materials Engineering, University of Alberta, Alberta T6G 1H9, Canada}

\title[An \textsf{achemso} demo]
  {Size Effect on Reaction Rate of Surface Nanodroplets}

\abbreviations{OA,OTS}

\begin{document}

\begin{tocentry}
\centering
 \includegraphics[height=4.5cm]{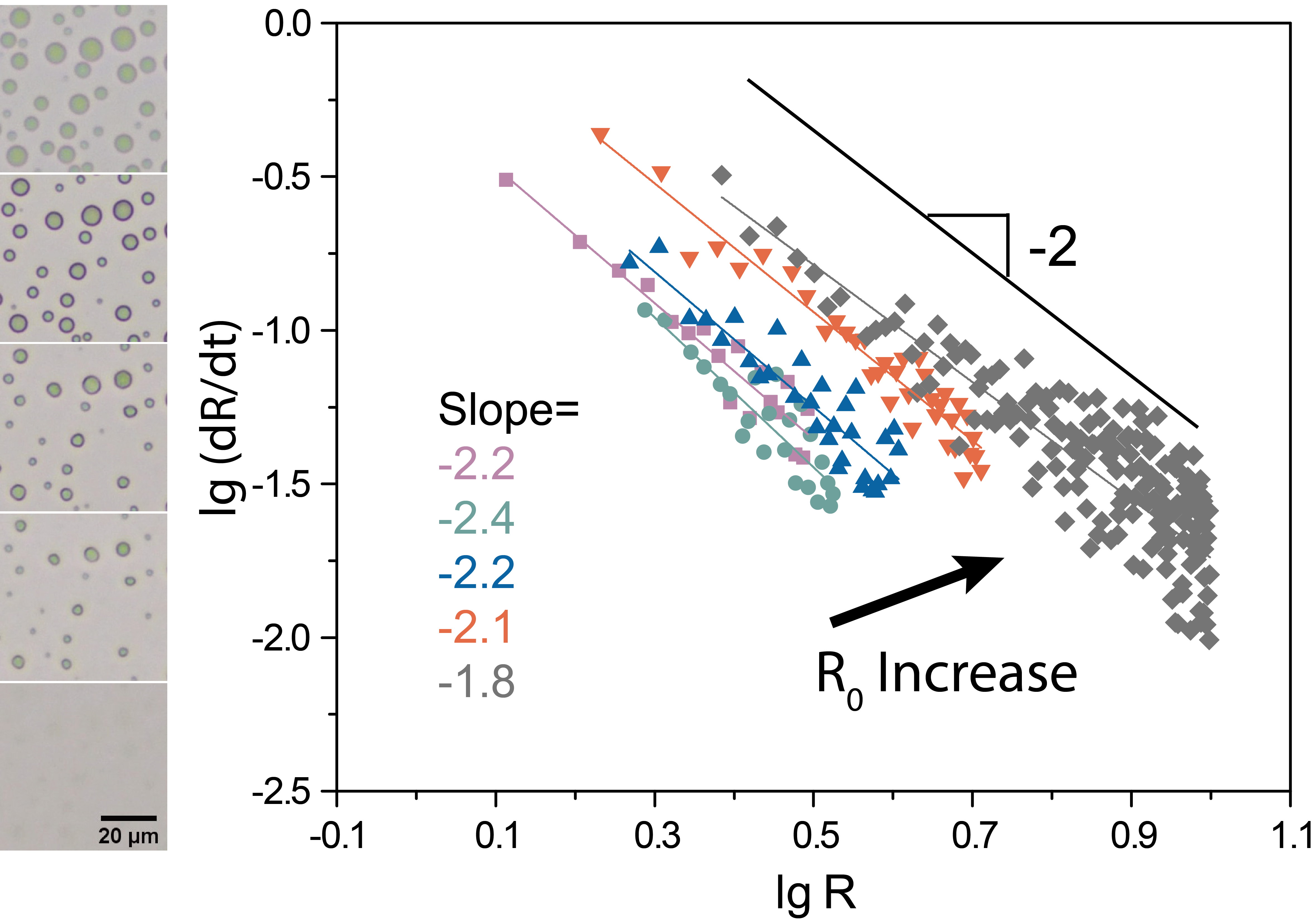}
\end{tocentry}

\begin{abstract}
Compartmentalizing reagents within small droplets is promising for highly efficient conversion and simplified procedures in many biphasic chemical reactions.
In this work, surface nanodroplets (i.e., less than 100 nm in their maximal height) were employed to quantitatively understand the size effect on the chemical reaction rate of droplets.
In our systems, a surface-active reactant in pure or binary nanodroplets reacted with the reactant in the bulk flow.
Meanwhile, the product was removed from the droplet surface.
The shrinkage rate of the nanodroplets was characterized by analyzing the lateral size as a function of time, where the droplet size was solely determined by chemical reaction rate at a given flow condition for the transport of the reactant and the product.
We found that the overall kinetics increases rapidly with the decrease of droplets lateral radius $R$, as $dR/dt \sim R^{-2}$.
The faster increase in the concentration of the product in smaller droplets contributes to accelerating reaction kinetics.
The enhancement of reaction rates from small droplet sizes was further confirmed when a non-reactive compound presented inside the droplets without reducing the concentrations of the reactant and the product on the droplet surface.
The results of our study improve the understanding of chemical kinetics with droplets.
Our findings highlight the effectiveness of small droplets for the design and control of enhanced chemical reactions in a broad range of applications.
\end{abstract}




\section{Introduction}
In-droplet chemistry is important for a wide range of reactions that involve reactants or products in immersible solvents. The large surface-to-volume ratio of microdroplets is desirable for efficient mass transfer crossing the interface in biphasic reactions \cite{kelly2007miniaturizing}. In heterogeneous catalytic reactions, catalyst adsorbed at the droplet surface has access to the compounds dissolved in both aqueous and organic phases \cite{wang2019ionophore,zhang2016compartmentalized}. Thanks to these advantageous features, droplet reactions are increasingly employed in drug discovery, synthesis of fine chemicals, biofuel conversion \cite{holland2020mass,yang2015compartmentalization}, and in fabrication of structured materials \cite{zhang2017pickering,liu2018surfactant}, such as polymeric microcapsules or microspheres \cite{wang2018polymerization}, porous membranes \cite{chinyerenwa2018structure}, lightweight materials \cite{tu2018ultralight}, and food grade foams \cite{su2020highly}. In tandem with fast analytic techniques, droplet reactions are indispensable in personalized medicine \cite{chen2021tumor,kang2014droplet}, point-of-care diagnostics \cite{kanitthamniyom2019magnetic}, anti-counterfeiting, \cite{wei2020integrated} food safety \cite{rothrock2013quantification}, or environmental monitoring \cite{baraban2011millifluidic}.
However, up to now what determine the rate of droplet reactions is not fully understood yet. 

Rates of droplet reactions have been reported to be dramatically different from their macroscopic counterparts. 
Cooks and Zare investigated a range of chemical reactions in microdroplets on flight produced by atomized sprays \cite{lee2015microdroplet,bain2016accelerated,lai2018microdroplets}. They found that Pomeranz-Fritsch synthesis of isoquinoline was much faster, sometimes even by a factor of $10^{6}$ faster than the same synthesis in bulk \cite{banerjee2017pomeranz}. Reactions that are extremely slow in bulk may occur readily in droplets. Notably, biomolecules including sugar phosphate and ribonucleoside can be synthesized at enhanced reaction rate in these flying droplets \cite{nam2017abiotic,nam2018abiotic}. Accelerated reaction kinetics was also observed in droplets from vapor condensation \cite{lee2020condensing}, and immersed droplets with liquid-liquid interface \cite{fallah2014enhanced,dyett2020accelerated}. The reactions are hydroperoxide generation \cite{lee2020condensing}, Mannich reaction \cite{fallah2014enhanced}, and dehydrocoupling of hydrosilanes \cite{dyett2020accelerated}. Apart from accelerated rates, droplet reactions may even produce new intermediates and product \cite{nam2017abiotic,baffou2014super}. For instance, new intermediates not present in bulk reaction was observed in droplets for phosphorylation between phosphoric acid and sugars \cite{nam2017abiotic}.

Several hypotheses have been proposed to account for the accelerated kinetics of droplet reactions. For droplets on flight, rapid solvent evaporation leads to droplet shrinkage in volume and consequently the increase in the reagent concentration, which may explain the enhanced reaction rate in the droplets \cite{marsh2019reaction,lee2015acceleration}.
Besides, a general effect of surface enrichment and orientation alignment for water-soluble probes, which results in a local electric field at the interface between water and a hydrophobic medium, might account in part for unique properties of chemical reactions in microdroplets \cite{xiong2020strongE,xiong2020strongC}.
In case of reagents localized at the liquid-liquid interface of the droplet, higher surface-to-volume ratio of the droplet contributes to enhanced mass transport of reagents and product throughout the interface, and thus the overall reaction kinetics \cite{fallah2014enhanced,dyett2020accelerated,li2020speeding}.
Other important effects are from the accelerated electron transfer and molecular configuration of chemicals at the interface with asymmetric environment on each side \cite{nam2017abiotic,nakatani1995direct}. The interfacial molecules are activated from solvation with lower energy barrier for the reaction and faster kinetics \cite{nakatani1995droplet,nakatani1996electrochemical}.
Though multiple possible mechanisms were proposed to interpret the accelerated kinetics, however, the quantitative analysis based on tracing in-situ droplet reaction is still missing.

Surface nanodroplets serve as an ideal model system for in-droplet chemistry to quantitatively study the reaction rate. Surface nanodroplets are droplets on a solid substrate in contact with a liquid immiscible with the droplet liquid \cite{lohse2015surface}. Here nano- refers to the height of the droplets less than 100 nm. The base diameter of the droplets ranges from hundreds nanometer to tens of micrometer and the volume from atto- to femto- litters. The size distribution and chemical composition of surface nanodroplets can be well controlled by a simple method called solvent exchange \cite{zhang2015formation,lu2015solvent,lu2016influence}. Surface nanodroplets are stationary, due to pinning on their boundary. Stability of surface nanodroplets enables us to follow in-situ the change in the droplet size resulted from chemical reactions \cite{lohse2015surface}, and to quantify the chemical composition in nanodroplets by sensitive molecular spectroscope, such as attenuated total internal reflection infrared \cite{dyett2018extraordinary} or surface enhanced Raman spectra \cite{li2019functional}. In particular, the stability of surface nanodroplets allows for supplying the reactant in an external flow under well-controlled conditions to investigate the effect of diffusion and convection on the reaction rate of droplets.

As a nanodroplet is usually surrounded by many neighbours, the collective effect from the reaction of neighbouring droplets is an important aspect to understand the rate of droplet reaction.
In our recent work, we studied the effects from external flow conditions on chemical kinetics of the biphasic reaction. In a model system, surface nanodroplets of oleic acid reacted with alkali dissolved in an external flow \cite{li2020speeding}.
The flow rate influenced the transport of both the reactant and the product. We obtained a scaling law of droplet shrinkage rate with the dimensionless number Peclet \cite{li2020speeding}. The reaction of neighbouring droplets hinders the droplet reaction, due to product diffused in the surrounding.  However, it remains unclear what the correlation is between the reaction rate and the droplet size and how a non-reactive component in droplets influences the reaction rate. 

In this work, we will focus on the dependence of reaction rates on the size of individual droplet with or without neighbouring effect.
The reactive droplets consist of either pure or binary liquid (binary droplet).
We are able to establish quantitative correlation between the reaction rate and the size of both pure and composite droplets.
The findings in our work may not only serve as the basis for improving the understanding of the altered kinetics of chemical reactions at the droplet surface, but also help to design the droplet reactions, heterogeneous catalysis, advanced materials fabrication, and other applications based on reactions with droplets.

\section{Methodology}
\subsection{Chemicals and materials}
Oleic acid (OA in brief, $\geq$90\%, Fisher), ethanol ($\geq$89\%, Fisher), sodium hydroxide ($\geq$97\%, Alfa Aesar), decane ($\geq$99\%, Fisher), and octadecyltrichlorosilane (OTS) ($\geq$95\% Fisher) were used as received without any further purification.
Ultra-pure Water (18.2 M$\Omega$cm, Millipore) was used in all experiments.
Silicon substrates (100 mm in diameter and 500 $\mu$m in thickness, Universitywafer, US) were hydrophobized with OTS by following a reported procedure \cite{lessel2015}.
Before use, coated substrates were sonicated in ethanol and then in water for 5 minutes, and dried in a stream of air.

\begin{figure}[h]
\centering
 \includegraphics[height=7.6cm]{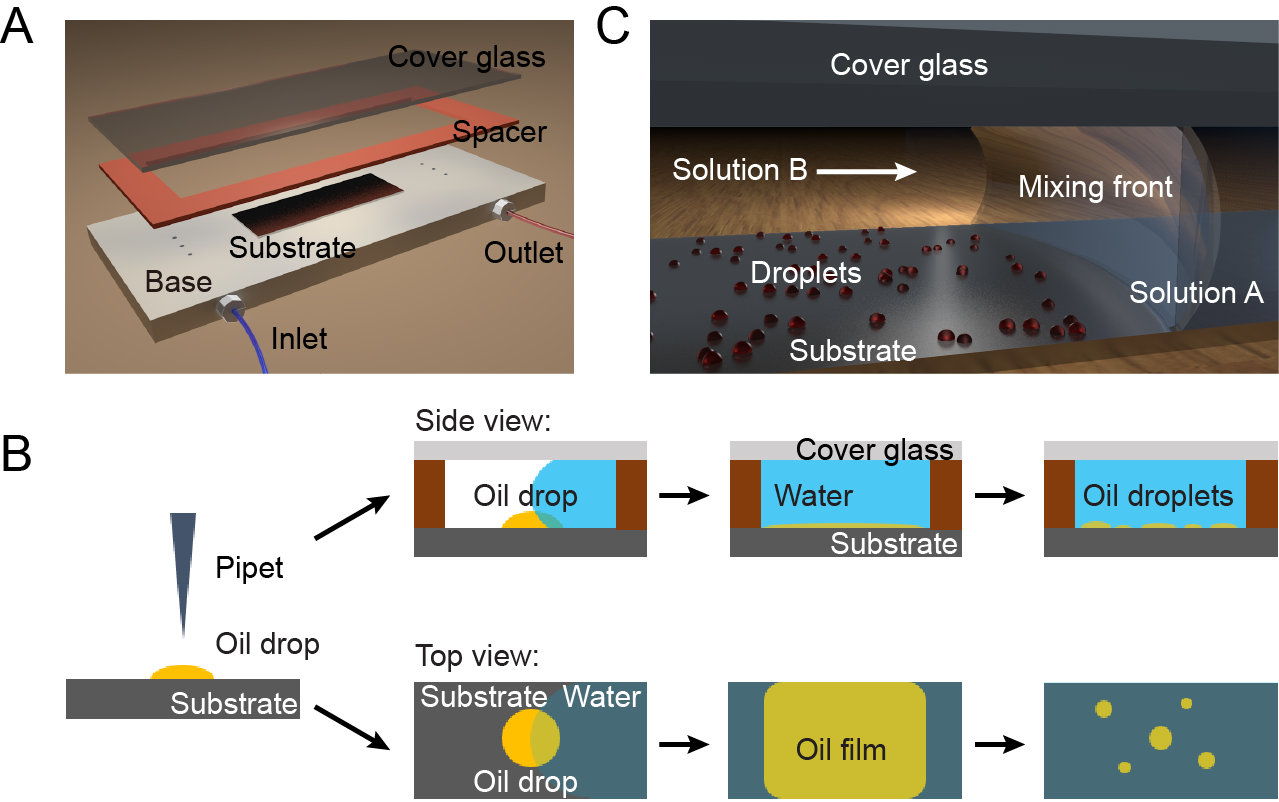}
  \caption{Sketch of the experiment setup: (A) The micro-channel for formation and reaction of surface nanodroplets.
  (B) Procedure of droplet formation by the method of film breaking.
  (C) Illustration of droplets formed by solvent exchange.
  The white arrow in the chamber points the direction of the alkali flow.}
\label{fgr:Setup1}
\end{figure}

A fluid channel was self-designed for droplet formation and reaction.
As sketched in Figure \ref{fgr:Setup1}A, the channel consisted of a polymethyl methacrylate (PMMA) base, an inlet and an outlet drilled in the base, a top cover glass, and a silicone spacer.
The micro-channel inside PMMA was drilled by hole puncher.
Figure S1 demonstrates the detailed layout of the PMMA base.
The thickness of the spacer was adjusted to control the channel height.
The substrate was located on the center of the base.
All experiments of droplet formation and chemical reaction were performed at temperature of 21$^\circ$C.

\subsection{Preparation of pure OA droplets with low surface coverage}
OA droplets with low surface coverage were prepared by the film breaking method \cite{qian2020one}.
Figure \ref{fgr:Setup1}B schematically demonstrates the film breaking method.
An OA drop was placed on the substrate (25 mm by 8 mm) by using a micropipette.
The initial OA drop was controlled at 500 $\mu$m in base radius. 
Then, the channel was assembled and sealed with spacers and clips.
The height $h$ and the width $w$ of the channel were respectively 2.4 mm and 15 mm.

After the channel was assembled, water was slowly injected into the channel through a syringe.
When the water-air interface passed the drop, the drop deformed into a film of oil.
The oil film broke up and formed several smaller OA droplets. 
The lateral diameters of the newly formed droplets ranged from several microns to tens of microns. 
After film breaking, around 10 to 30 droplets were formed in each experiment.
The distances between newly formed droplets were from tens of micrometers to several millimeters.

\subsection{Preparation of pure OA droplets with high surface coverage}
OA droplets with high surface coverage were formed by solvent exchange \cite{zhang2015formation}, to test the collective effect from neighbouring droplets on the reaction kinetics.
In solvent exchange experiments, the height $h$ and the width $w$ of the channel were respectively 0.4 mm and 15 mm.
Before solvent exchange, two solutions were prepared.
Solution A was 2.6\% (v/v) OA in the mixture of ethanol and water with the ratio of 7:3 (v/v).
At first, the fluid channel was filled with solution A.
Water (solution B) was then injected into the channel with a programmable syringe pump (New Era, NE-1000).
The volume flow rate of solution B was 500 $\mu$L/min.

When the mixing front of two solutions passes over the substrate, as sketched in Figure \ref{fgr:Setup1}C, droplets nucleate and grow due to the oversaturation of OA in the mixing front.
In each group of the experiments, around 4,000 surface droplets can be formed in the field of view (0.34 mm$^2$).
Thanks to the well-controlled fluid conditions, droplets for different experiments had consistent size distribution and number density.
The initial surface coverage $SC_0$ were $\approx$ 25\%. $SC_0$ is defined as the proportion of the substrate surface occupied by droplets after droplet formation.
The averaged droplet radius was controlled at $\sim$ 2.4 $\mu$m.

\subsection{Preparation of binary droplets}
Binary droplets of OA and decane with low surface coverage were also prepared by film breaking method.
A binary drop of 5\%/10\% decane and 95\%/90\% OA in volume was placed the substrate.
Water was slowly introduced into the chamber by a syringe, and the initial drop broke up to several smaller droplets.
Binary droplets with high surface coverage were prepared by solvent exchange method.
Solution A was prepared by adding decane and OA into 70\% ethanol aqueous solution.
Then standard solvent exchange was performed.

\begin{figure}[h]
 \centering
 \includegraphics[height=5cm]{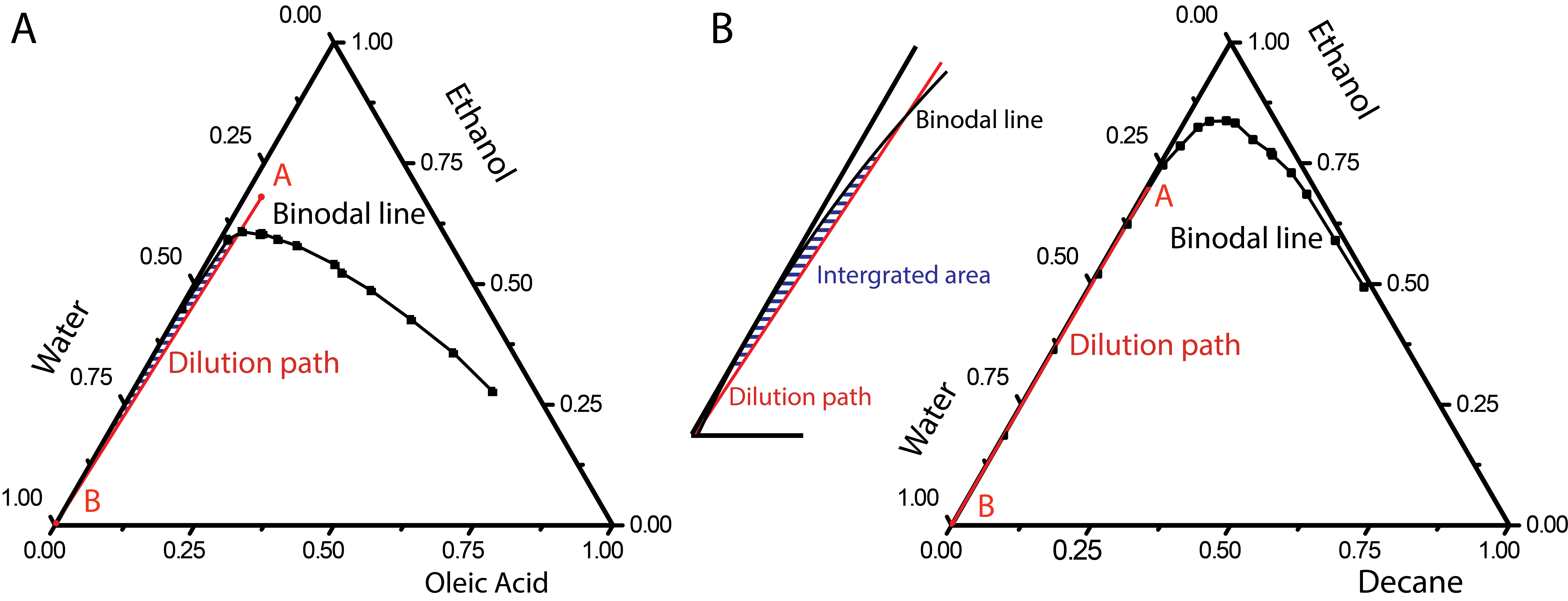}
  \caption{Ternary phase diagram of (A) oleic acid and (B) decane. Points A and B in plots are the composition points of solutions A and B. The shaded area surrounded by the dilution curve (red line) and binodal line (black line) in the plot reflect the oversaturation level of OA (A) and decane (B) in the mixing front.}
\label{fgr:Phase}
\end{figure}

When producing binary droplets by solvent exchange method, the composition of binary droplets was determined by the oversaturation level of each component in the mixing front \cite{li2018formation}.
The oversaturation level can be expressed on the ternary phase diagram, by the area surrounded by the binodal curves and the dilution path.
Based on the solution composition in initial solutions and ternary phase diagram, the corresponding composition of surface droplets can be approximately calculated.
The details of the calculation can be found in literature \cite{lu2015solvent,lu2016influence}.
Nine points in the ternary phase diagrams of oleic acid and decane in Figures \ref{fgr:Phase}A\&B were taken from IUPAC-NIST Solubility Database and additional points were obtained by titration for a complete diagram.
Table \ref{tbl:binary} lists all solution A compositions, solution B flow rates, and the composition of formed surface nanodroplets.

\begin{table*}
\small
  \caption{\ Solvent exchange conditions and corresponding droplet compositions}
  \label{tbl:binary}
  \begin{tabular*}{\textwidth}{@{\extracolsep{\fill}}cccc}
    \hline
    OA in solutin A & Decane in solution A & Solution B flow rate & Decane in droplet\\
    (v/v) & (v/v) & ($\mu$L/min) & (v/v)\\
    \hline
    2.8\% & 0\% & 500 & 0\%\\
    2.8\% & 0.16\% & 500 & 3\%\\
    2.8\% & 0.21\% & 500 & 5\%\\
    2.8\% & 0.25\% & 500 & 7\%\\
    2.8\% & 0.28\% & 500 & 9\%\\
    \hline
  \end{tabular*}
\end{table*}

After solvent exchange, around 5,000 binary droplets were produced with a similar number density and size distribution in the field of view.
The initial surface coverage of binary droplets was controlled at $\approx$ 34\% for all groups of experiments.
We note that the group with 10\% decane in droplets was an exception.
Due to the coalescence of droplets during the solvent exchange, only around 3,000 droplets formed in the recorded area, and the surface coverage was around 38\%.

\subsection{Flow and solution conditions for droplet reaction}
An aqueous solution (solution C) of sodium hydroxide was prepared to react with OA droplets.
The concentration of NaOH was constant at $4.0\times10^{-4}$ M for all experiments.
We chose the specific concentration of NaOH for a suitable reaction rate that the reaction process is not too fast or too slow to be followed in-situ.

After droplet formation, solution C was injected into the channel by the syringe pump at a constant flow rate.
In experiments of droplets with low surface coverage, the flow rate in volume $Q$ were controlled at 167 $\mu$L/min, where the corresponding Peclet number $Pe$ was 52.
$Pe$ is defined as $Pe=Q/(wD)$, where $Q$ and $w$ are the volume flow rate and the channel width.
In experiments of pure droplets with high surface coverage, the flow rates $Q$ varied from 100 to 645 $\mu$L/min, where corresponding Peclet numbers were $Pe=$15-198.
In experiments of binary droplets with high surface coverage, $Q=$167 $\mu$L/min and $Pe=$52.

Reacting droplets were recorded by an upright microscope (Nikon H600L, 10x objective lens) with a video camera (Nikon, DS-Fi3, 0.24 $\mu$m/pixel, 5.0 fps).
A white-light LED lamp was used for bright field imaging.
How droplets were selected and presented in different sections was discussed in Supporting Information.
ImageJ and self-written Matlab codes were used to process and analyze the filmed images. 
The base area, $A$, of droplets as function of time, and the surface coverage, $SC$ were extracted frame by frame.
Time zero ($t_0$) was defined as the moment when solution C reached droplets.
$t_0$ was separately predetermined by flowing the alkali solution at the same flow rate in an empty fluid channel.

We assume that the droplets are spherical caps. Their boundaries are expected to be circular, due to low surface tension of the surfactant product and low pinning effect on the homogeneous substrate.
Only occasionally droplets were seen to be pinned, possibly due to imperfection of the substrate.
The base radius of the droplet $R$ was calculated as $R=(A/\pi)^{0.5}$.
Droplets that are too large ($R_0>$ 13 $\mu$m) or too small ($R_0<$ 3 $\mu$m) in their initial sizes were excluded, to avoid too long or too short reaction time.

The experimental uncertainty was estimated to be the spatial resolution of the microscope (0.24 $\mu$m, pixel size).
We note that although there may be systematic errors in the dimension due to the spatial resolution limitation in our optical images, such systematic errors do not influence our results determined by the change of droplet size.

\section{Results and discussion}
The reaction between oleic acid (OA) droplets and alkali in the solution is shown below
\begin{equation}
\label{eq:Reation}
 \begin{aligned}
  CH_{3}(CH_{2})_{7}CHCH(CH_{2})_{7}COOH + OH^{-} \\
  \rightleftharpoons CH_{3}(CH_{2})_{7}CHCH(CH_{2})_{7}COO^{-} + H_{2}O.
 \end{aligned}
\end{equation}

\begin{figure}[h]
\centering
 \includegraphics[height=5cm]{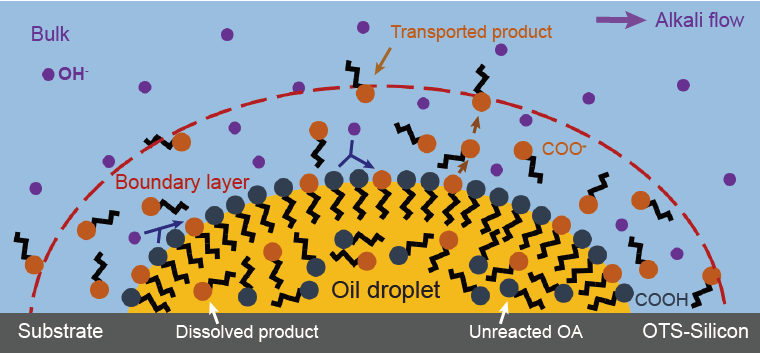}
  \caption{Reaction between alkali in the flow and an OA droplet.
  The –COOH head group (indigo) reacts with OH$^{-}$ (purple) from bulk, producing –COO$^{-}$ head group (orange).
  Black chain in the figure is the hydrophobic tail of reactant and product.
  The reaction is indicated by the blue arrow.
  The red dashed line show the diffusive boundary layer of product out of the droplet.
  Purple arrow on the top indicate the direction of the flow.}
\label{fgr:Setup2}
\end{figure}

Figure \ref{fgr:Setup2} sketches the process of the droplet reaction.
Carboxyl groups of OA tended to stay at the waterside of the interface, since they are hydrophilic.
Hydroxide ions in the solution were transported to the droplet by the flow, and reacted with the carboxyl groups.
OA is converted to oleate.
After reaction, oleate dissolved in the OA droplet and the surrounding solution.
The later and was carried away by the flow.
The overall process consisted of (1) mass transport of alkali from the laminar flow to droplet interface, (2) reaction between alkali and carboxyl at the droplet interface, (3) product (oleate) desorption from the droplet surface, and (4) product dissolution in the droplet and transport by the flow \cite{li2020speeding}.

\subsection{Size effect on reaction kinetics of pure droplets}
\begin{figure}[h]
 \centering
 \includegraphics[height=6cm]{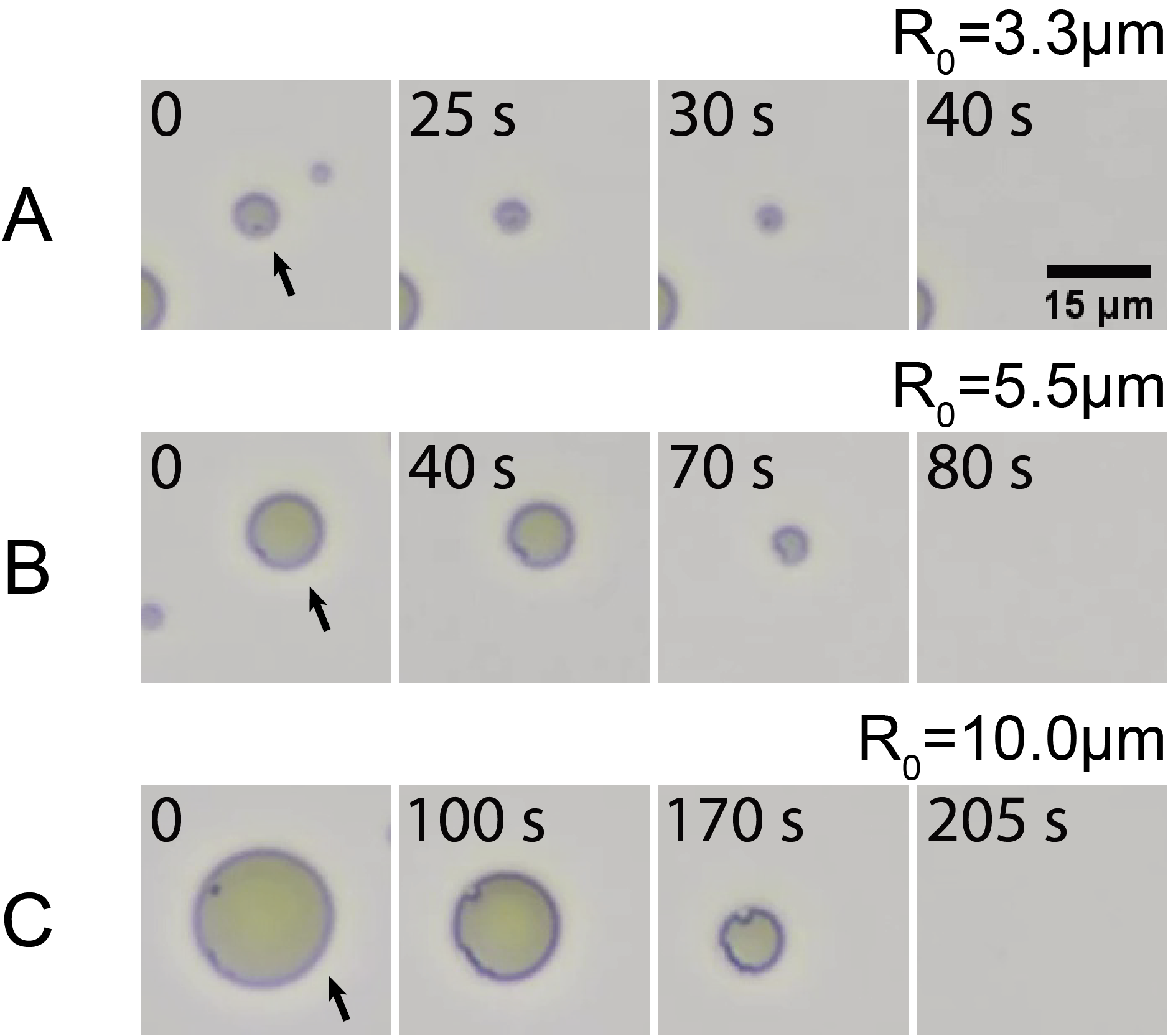}
 \caption{Optical images of surface nanodroplets reacting with the alkali solution. The initial radius at $t=0$ of droplets (pointed by the arrow) in (A)-(C) were $R_{0}=3.3$ $\mu$m, 5.5 $\mu$m, and 10.0 $\mu$m, respectively. The black arrow points to the reacting droplet.}
 \label{fgr:Film1}
\end{figure}

Three series of snapshots in Figures \ref{fgr:Film1}A-C reveals progressive shrinkage of reacting droplets with different initial sizes.
The second and third columns of snapshots in Figure \ref{fgr:Film1}A-C reflect the continuous shrinkage in the lateral radius of the droplet from the moment when reactant solution reached the droplets till disappearance in the fourth column.

\begin{figure}[h]
 \centering
 \includegraphics[height=4.5cm]{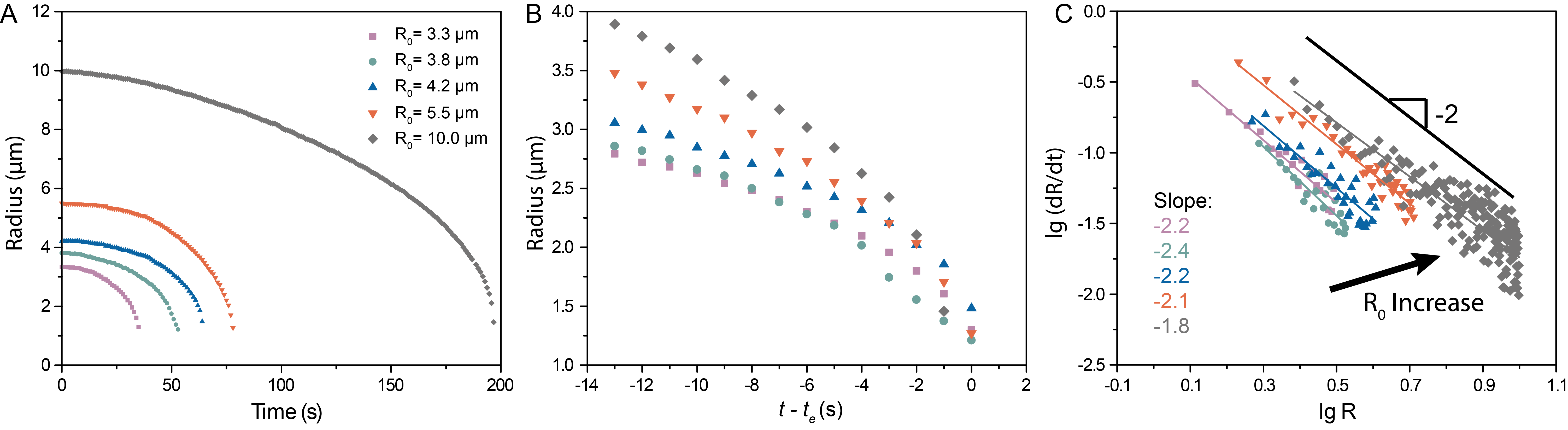}
 \caption{(A) Lateral radius $R$ of OA droplets with different initial sizes as a function of time. $R_0$ represents the initial lateral radius at $t_0$.
 Legends in (A) carry the same meaning throughout Figure \ref{fgr:Film2}.
 (B) Shrinkage of each droplet in the last 14 seconds when they are larger than 1.2 $\mu$m rescaled from (A).
 (C) Dissolution rate $\dot R$ as function of $R$ of droplets with different $R_0$ values.
 The coefficient $R^2$ of the fitting ranges from 0.74 to 0.93.
 The black solid line represent the result $\dot R\propto$ $R^{-2}$ from the scaling analysis.
 At least three experimental replicates were performed under the same conditions in this section.
 Overall, 12 curves were obtained to confirm the correlation in Figure \ref{fgr:Film2}C.}
 \label{fgr:Film2}
\end{figure}

The lateral radius of droplets was analyzed as the function of time and the results are plotted in Figure \ref{fgr:Film2}A.
$R_0$ represents the initial lateral radius of droplets at $t_0$.
In general, the shrinkage of droplets speeds up as time proceeds.
In other words, the rate of the reaction is dependent on the size of the reacting droplet.

The dissolution curves of droplets near the end were plotted in Figure \ref{fgr:Film2}B.
The time zero here is re-scaled by subtracting $t_e$ from time, where $t_e$ is the last second when the droplet was larger than 1.2 $\mu$m in lateral radius.
Figure \ref{fgr:Film2}B reveals that the droplets tend to have similar dissolution rates when they are similar in the instantaneous lateral radius.
These results suggest that the instantaneous lateral radius of the reacting droplets plays the major role in determining dissolution rate.

We compared the dissolution rates of the same droplet at different times to understand the effects from the droplet size (i.e., radius) on the surface kinetics quantitatively.
The average dissolution rate $\dot R$ in each second was determined by calculating the variation of lateral radius $R$.
In Figure \ref{fgr:Film2}C, the dissolution rates $\dot R$ were plotted as function of the instantaneous radius $R$.
All the droplets collapse into the universal scaling with the exponent of -1.9 to -2.2, while the pre-factor varies with the initial radius.

Our results suggest that the droplet dissolution rate was accelerated by reducing the droplet radius $R$, scaling with $R^{-2}$.
Although entirely different in the mechanism, this dependence of the dissolution rate on $R^{-n}$ is close to earlier reports in which acceleration in reaction rate was attributed to the thermodynamic variation from chemical adsorption and desorption at the interface.\cite{fallah2014enhanced,dyett2020accelerated}

\subsection{Reaction kinetics of droplets at high surface coverage}
\begin{figure}[h]
\centering
  \includegraphics[height=9.5cm]{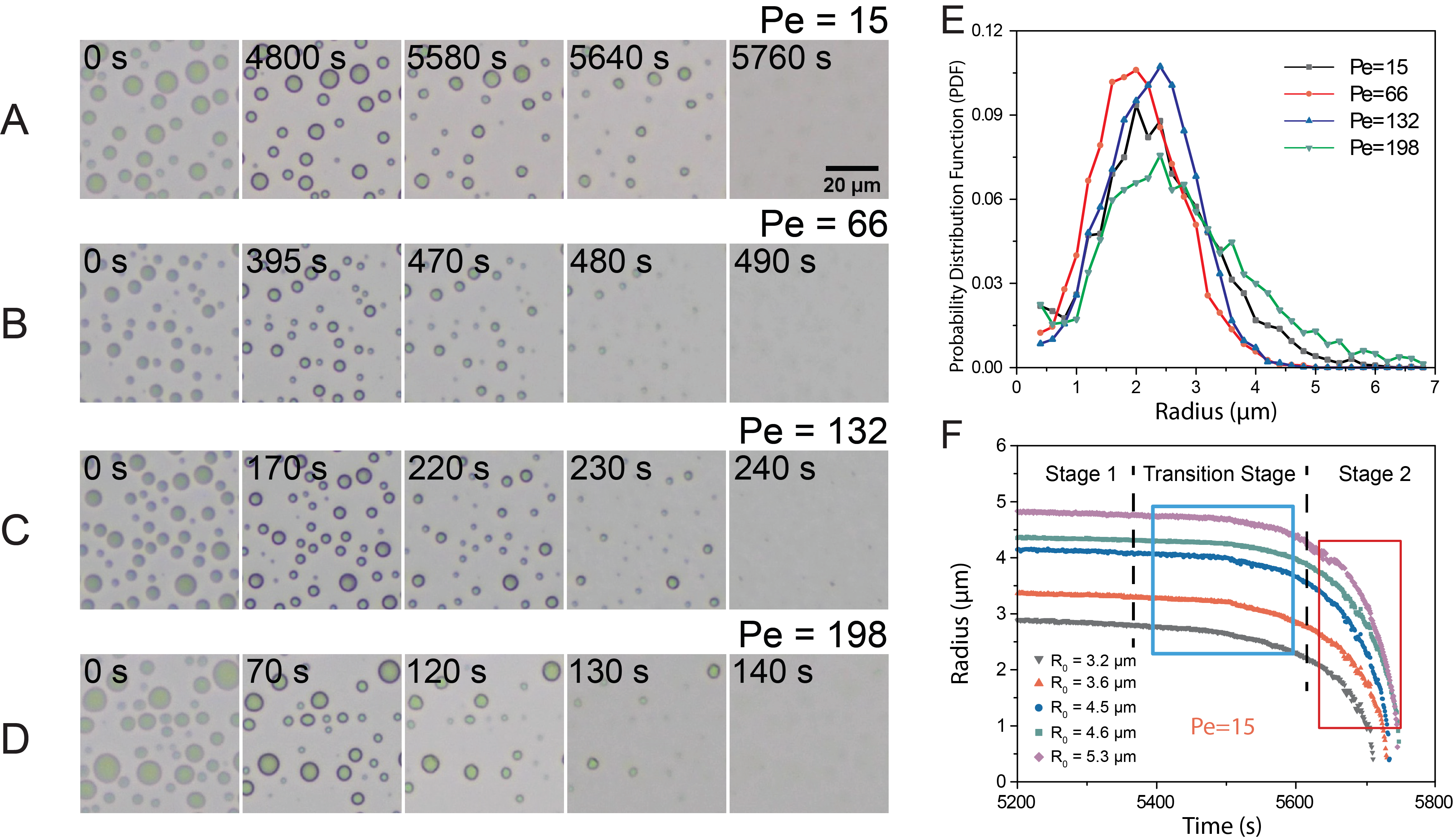}
  \caption{(A)-(D) Screenshots of nanodroplets with high surface coverage shrinking in the alkali flow at different flow rates.
  The flow rates were controlled at 50, 215, 430 and 645 $\mu$L/min from (A) to (D), which corresponds to Peclet numbers from 15 through 198.
  (E) Probability distribution functions (PDF) of droplet base radius after droplet formation.
  Only droplets with the base radius larger than 0.3 $\mu$m are demonstrated in (E).
  (F) Lateral radius $R$ of droplets with different initial sizes as function of time from $t=5200$ s to the end of the dissolution.
  Data in (F) were obtained from the analysis of the video of \ref{fgr:Exchange1}(A) ($Pe=15$).
  Black dashed lines in (F) divide dissolution stages.}
  \label{fgr:Exchange1}
\end{figure}

Solvent exchange was applied to produce OA droplets with a large number density.
The initial surface coverage of OA droplets was fixed at $\approx$ 25\%.
The probability distribution functions (PDF) of size of OA droplets in Fig \ref{fgr:Exchange1}E suggests that the size distributions of droplets were all similar at $t_0$ in the four groups of experiments. 
Series of snapshots in Figures \ref{fgr:Exchange1}A-D reveals that OA droplets progressively shrank at different flow rate of the reactant solution.
It is noteworthy that the droplets did not shrink from $t_0$, but remained at almost its initial radius for certain time.
This period is defined as stage 1, in which the product from upstream droplets transported along the flow hindered the downstream reactions.
It is a different trend when compared to the low surface coverage cases, where the hindering effect was relatively weak, and the dissolution starts immediately after the alkali solution entering the field.
As expected, the duration of stage 1 was shorter at higher flow rate of the reactant solution.

Figure \ref{fgr:Exchange1}F presented the lateral radius of droplets as a function of time after stage 1. 
Data in Figure \ref{fgr:Exchange1}F were extracted from the analysis of the video of Figure \ref{fgr:Exchange1}A ($Pe=15$).
After stage 1, dissolution rate of droplets continuously increased until droplets finally disappeared.
At stage 2, most upstream droplets were depleted and removed by the flow, and the surface coverage of nearby droplets also decreased.
The hindering effect does not play a significant role anymore at stage 2.
We also defined a transition stage between stages 1 and 2, in which droplets started to dissolve at an observable rate, while the hindering effect from products was still pronounced.

\begin{figure}[h]
\centering
  \includegraphics[height=4.5cm]{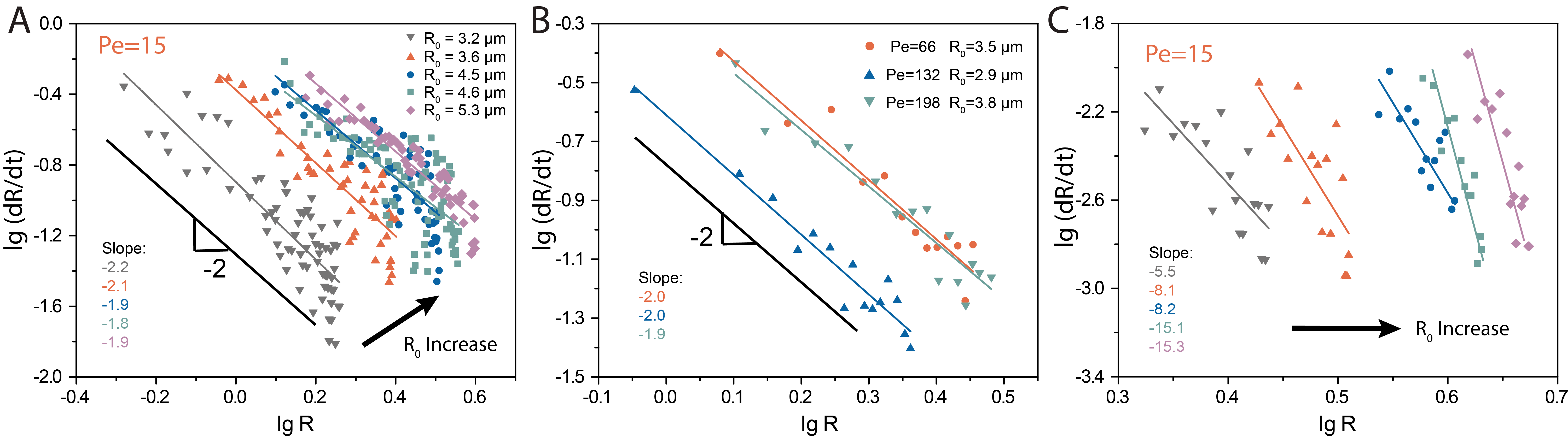}
  \caption{(A) Scaling relationship between $\dot R$ and $R$ for droplets with different initial droplet sizes in stage 2 ($Pe=$ 15).
  Data in (A) were obtained from the red box in Figure \ref{fgr:Exchange1}F.
  Legends in (A) carry the same meaning in (C).
  (B) Scaling relationship between $\dot R$ and $R$ at the higher Peclet numbers ($Pe=$ 66, 132, and 198) in stage 2.
  The coefficient $R^2$ of the fitting ranges from 0.49 to 0.93.
  The black solid lines in (A)\&(B) represent the result $\dot R\propto$ $R^{-2}$ from the scaling analysis.
  (C) Scaling relationship between $\dot R$ and $R$ for droplets with different initial sizes at the transition stage ($Pe=$ 15).
  Data in (C) were obtained from the blue box in Figure \ref{fgr:Exchange1}(F).
  The coefficients $R^2$ of the fitting range from 0.55 to 0.79.
  For each Peclet number, more than 10 droplets were analyzed to establish the correlation in this section.}
  \label{fgr:Exchange2}
\end{figure}

Since droplets did not shrink much in the first stage, we at first focused on the second stage of the shrinkage.
Based on the data marked by the red box in Figure \ref{fgr:Exchange1}F, we analyzed the dissolution rate $\dot R$ in the second stage.
The surface coverage at $t=5,640$ $s$ was 7.1\%.
As shown in Figure \ref{fgr:Exchange2}A, the dissolution of individual droplets obeys the scaling law $\dot R \propto R^{-n}$ with $n=1.8-2.2$, which agrees well with that found for low surface coverage droplet reaction.
We also analyzed the higher Peclet numbers data.
Figure \ref{fgr:Exchange2}B shows the exponent $n=1.9-2.1$ at $Pe=$66, 132, and 198.
Data in Figure \ref{fgr:Exchange2}B were extracted from the analysis of the video of Figures \ref{fgr:Exchange1}B-D.
Only the result of one droplet was presented for each group of Peclet numbers.
The data from droplets with different initial sizes at the same Peclet number are presented in Figure S2.
When the surface coverage is relatively low, the size effect was more pronounced rather than the hindering effect.

We then analyzed the dissolution data in the transition stage ($t=$5,400 $s$ to 5,600 $s$, blue box in Figure \ref{fgr:Exchange1}C).
The surface coverage at $t=5,400$ $s$ was 14.6\%.
The droplets shrank slowly, likely due to the hindering effect triggered by the high surface coverage.
We calculated the average shrinkage rate every 5/10 seconds and obtained the scaling exponents (Figure \ref{fgr:Exchange2}C).
Compared with the results extracted in stage 2, two general features were: (1) results in transition stage usually have larger scaling exponents between $\dot R$ and $R$ compared with the stage 2, (2) droplets with larger initial sizes tend to have larger exponents $n$.

The larger scaling exponents in the transition stage may be explained by the attenuation of the hindering effect.
In the transition stage, the upstream droplets were gradually depleted.
The concentration of the products in the flow continuously decreased, and thus the hindering effect become weaker.
Consequently, the shrinkage rate, $\dot R$, increased with time.
Besides, due to the low shrinkage rate, the droplet size was reduced little and contributed less to the increasing shrinkage rate.
As a result, the scaling exponents became much higher than $-2$.
Besides, droplets with small initial sizes tend to have a large product transportation rate, suggesting that small droplets can immediately finish the transition stage \cite{li2020speeding}.
Thus, the scaling exponent for smaller droplets could be rather close to -2 than their larger counterparts.

\subsection{Size effect on reaction of binary droplets}
\begin{figure}[h]
\centering
  \includegraphics[height=8.8cm]{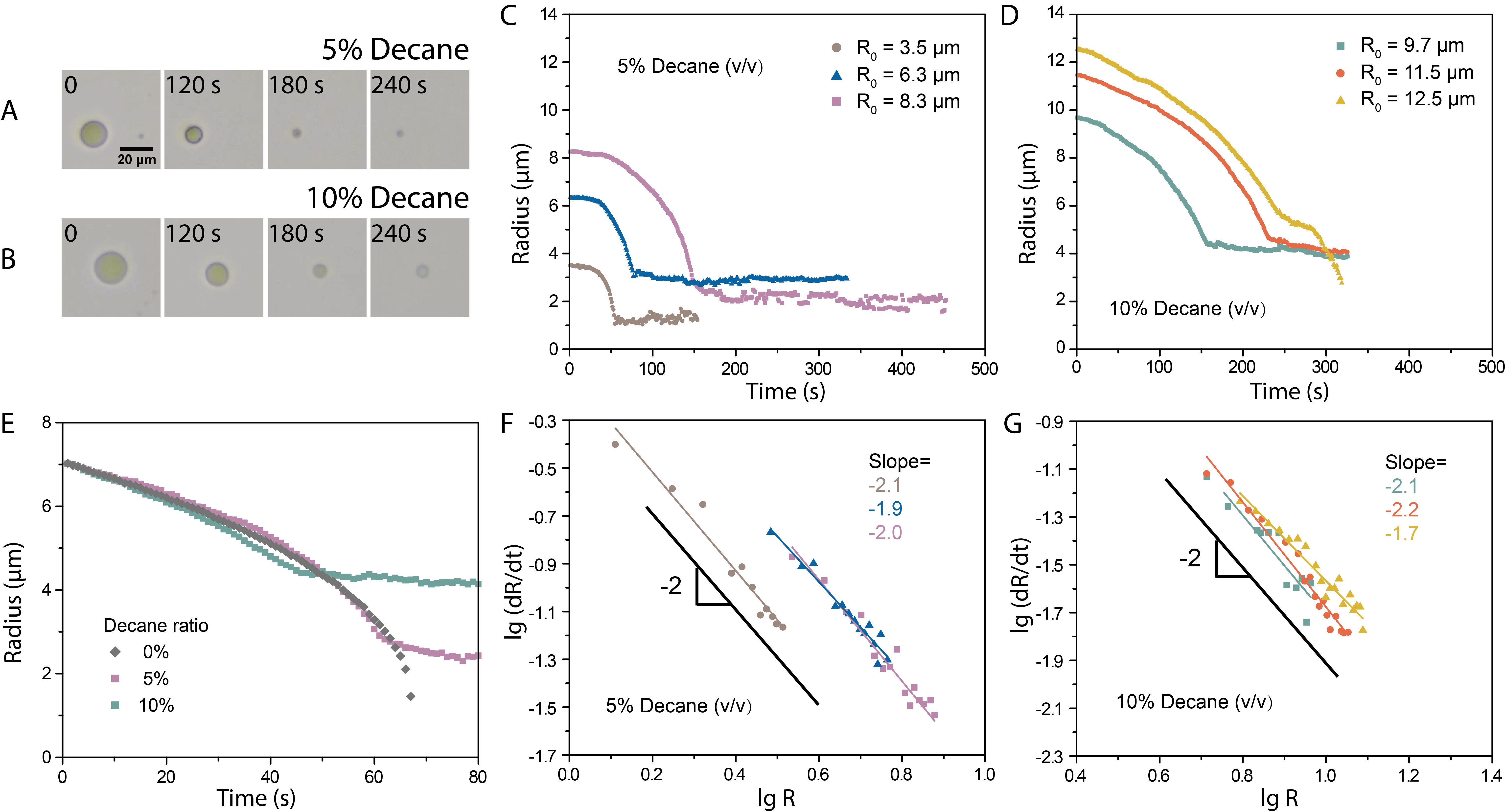}
  \caption{(A)\&(B) Optical images of decane/OA binary droplets reacting and dissolving in the alkali flow.
  The volume ratio of decane was 5\% in (A) and 10\% in (B).
  The corresponding Peclet number $Pe$ was 10.
  (C)\&(D) Lateral radius $R$ of binary droplets with decane volume ratio 5\%\ and 10\% as function of time.
  Legends in (C)\&(D) carry the same meaning throughout Figure \ref{fgr:Binary}.
  (E) Shrinkage of droplets of 0\% ($R_0=10.0$ $\mu$m), 5\% ($R_0=8.3$ $\mu$m), and 10\% ($R_0=9.7$ $\mu$m) decane ratio.
  The time zero is defined as the moment when lateral radius of each droplet decreased to $R=7.0$ $\mu$m.
  (F)\&(G) Dissolution rate $\dot R$ calculated every 10 seconds as function of $R$.
  Results in (F)\&(G) were from the analysis of data in (C)\&(D).
  The coefficient $R^2$ range from 0.71 to 0.97.
  The black solid lines in (F)\&(G) represent the result $\dot R\propto$ $R^{-2}$ from the scaling analysis.
  For each droplet composition, at least three experimental replicates were performed under the same conditions.
  Overall, 14 droplets were analyzed to obtain the correlation in Figure \ref{fgr:Binary}F\&G.}
  \label{fgr:Binary}
\end{figure}

To test how the addition of the non-reactive, surface inactive components will influence the biphasic reaction at droplet surface, binary droplets were formed with low surface coverage by film breaking method.
Compared with decane, OA is surface active, due to the carboxylic acid group.
Figures \ref{fgr:Binary}A\&B are snapshots of binary droplets shrinking in the alkali flow.
Assuming that the compositions of droplets were consistent with the compositions in the oil film, the decane ratio of droplets shown in Figures \ref{fgr:Binary}A\&B were 5\% and 10\% (v/v).
Similar to pure droplet reaction with low surface coverage, binary droplets immediately shrank when alkali entered the field of view.
Adding the non-reactive component, droplets did not completely disappear on the substrate at the end in contrast to OA droplets.
Non-reactive and insoluble decane were left on the substrate in the form of smaller droplets, as demonstrated by the fourth snapshot in Figures \ref{fgr:Binary}A\&B.
10\% decane droplets cause a larger residual droplet when compared to 5\% decane binary droplets with the similar initial sizes, as 10\% decane binary droplets contain more decane, which made up the main part of the residual droplets.

Figures \ref{fgr:Binary}C\&D are the lateral radius of binary droplets as function of time.
After the alkali reached binary droplets, the dissolution was gradually accelerated with the decreasing radius, and finally stopped when the OA component in binary droplets was run out.
Figure \ref{fgr:Binary}E compared the dissolution process of 5\% and 10\% decane binary droplets with the pure OA droplet, where the time zero was set as the moment when droplets dissolved to $R=7.0$ $\mu$m.
Importantly, the droplets dissolved at the same rate when they have consistent sizes, insensitive to the decane ratios.
This result suggests that the ratio of the non-reactive component does not influence the biphasic reaction in surface nanodroplets significantly. 
We believe that the surface inactive decane tends to stay at the inner part of the binary droplet, while surface-active OA mainly locates at the droplet outer layer.
The concentration of OA at the droplet surface and thus reaction rate is not influenced by the decane ratio in droplets.

The scaling analysis of the relationship between $\dot R$ and $R$ are shown in Figures \ref{fgr:Binary}F\&G.
The size effect ($\dot R \propto R^{-n}$, $n\sim2$) can also be found in binary droplets.
The addition of non-reactive components does not influence the size effect including the scaling exponent.
This result also suggests that the non-reactive decane with low surface activity was hidden in the inner part of the binary droplet. 
As a result, the $-2$ exponent can be maintained even the ratio of decane in binary droplets continuously increase as the reaction proceeds.

\subsection{Binary droplet reactions at high surface coverage}
\begin{figure}[h]
\centering
  \includegraphics[height=8.6cm]{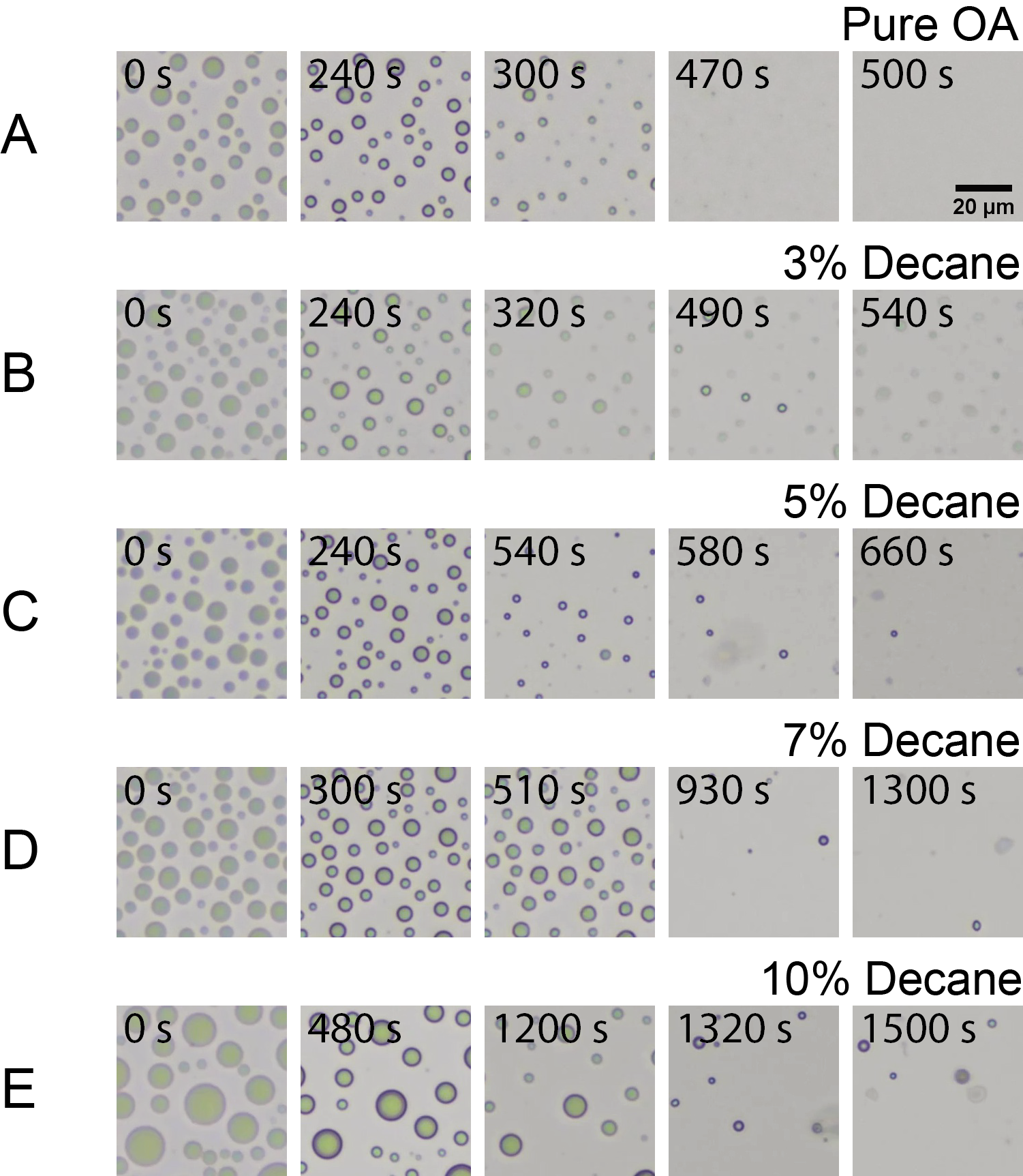}
  \caption{(A)-(E) Optical images of nanodroplets with different ratios of OA and decane reacting with the alkali flow.
  The approximate decane contents in surface droplets were from 0 to 10\% in (A) to (E) respectively.}
  \label{fgr:SC1}
\end{figure}

Binary droplets with high surface coverage were formed by solvent exchange and their reaction kinetics was investigated.
Figures \ref{fgr:SC1}B-E display snapshots of binary droplets with different decane compositions formed by solvent exchange, while the snapshots of pure OA (i.e., 0\% decane) droplets were arranged in Figure \ref{fgr:SC1}A for comparison purposes.
The amount of added non-reactive content increases as the concentration of decane increases, and the time needed to finish the dissolution process becomes longer.
We note that decane is insoluble and remains on the substrate even after finishing the dissolution process (see the fifth snapshots) as found in the case for binary droplets with low surface coverage.

\begin{figure}[h]
\centering
  \includegraphics[height=4.6cm]{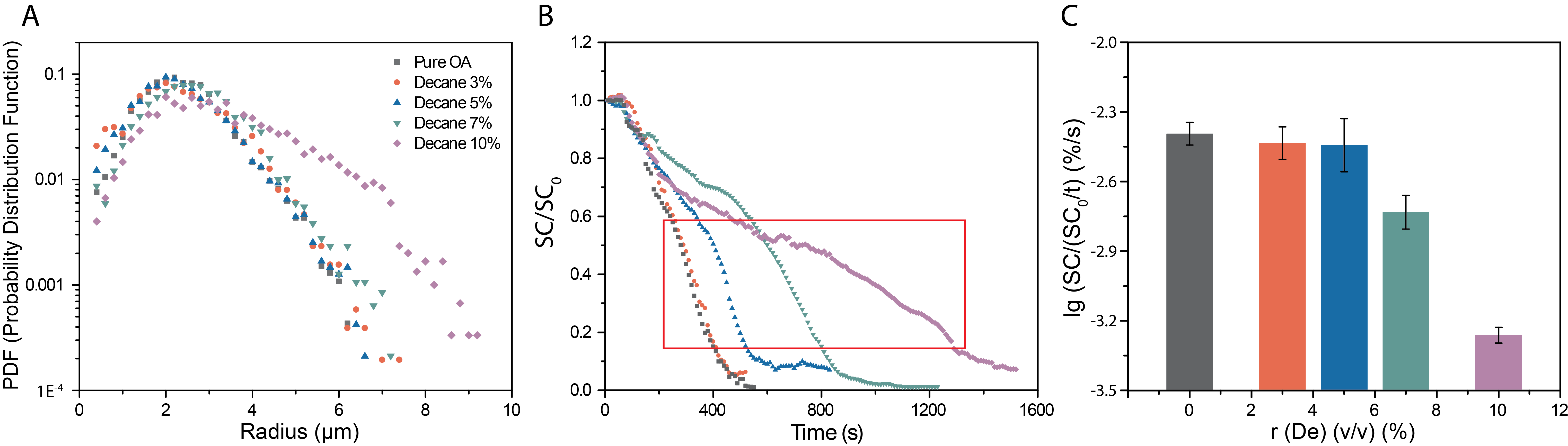}
  \caption{(A) Probability distribution functions (PDF) of droplet base radius. Only droplets with the base radius larger than 0.3 $\mu$m are demonstrated in (A). (B) At different flow rates, normalized surface coverage $SC/SC_0$ varied with time. $SC_0$ is the initial surface coverage. (C) Average dissolution rate at stage 2 in terms of normalized surface coverage (calculated from $SC/SC_{0}=$0.60 to $SC/SC_{0}=$0.15). $r$ is the volume ratio of decane in binary droplets.
  Error bars in (C) were calculated from three individual runs of experiments.}
  \label{fgr:SC2}
\end{figure}

Figure \ref{fgr:SC2}A shows the probability distribution functions (PDF) of binary droplets with different OA/decane ratio.
The initial droplet size distributions are similar to each other, except for the 10\% decane group.
(n\% decane refers to the decane composition in droplets.)
In the 10\% decane group, the number density of droplets decreased to around 3000, suggesting that some droplets merged and became larger, as confirmed in Figure \ref{fgr:SC1}E.
Consistently, 10\% decane group shows a relatively less population of droplets for the smaller lateral radius, while a much greater number of droplets were confirmed for the larger lateral radius, as presented in Figure \ref{fgr:SC2}A.

With the collective effect from high surface coverage of droplets, the diffusion of a binary droplet became complicated to be quantified without considering the neighbouring droplets. Hence, instead of presenting the lateral radius of a single droplet, we take a large number of droplets in a specific area as the entity to reduce variation from individual droplets surrounded by different neighbours.

The normalized surface coverage $SC/SC_{0}$ (normalized by the initial surface coverage after droplet formation) were analyzed quantitatively as a function of time in Figure \ref{fgr:SC2}B.
Binary droplets with the higher volume ratio of decane have different dissolution rate for each concentration, while 3\% decane binary droplets were similar to pure droplets.

Figure \ref{fgr:SC2}C shows the relationship between dissolution rate (see the marked region by the red box in Figure \ref{fgr:SC2}B) and decane volume ratio in droplets. 
The vertical and horizontal axes are respectively the normalized surface coverage and the decane volume ratios in droplets, which were estimated with the three-phase diagram.
Data in Figure \ref{fgr:SC2}C were calculated from $SC/SC_0$=0.15 to $SC/SC_0$=0.60 (see red box in Figure \ref{fgr:SC2}B). 
Notably, the dissolution rate at the fast dissolution stage (stage 2) of pure OA, 3\% decane and 5\% decane group were similar to each other. 
The existence of decane does not slow down the reaction rate in these groups significantly. 
In contrast, the 7\% decane group was a little slower, while the 10\% decane group was much slower than the other group. 
In the 10\% decane group, much larger droplet sizes contribute to decelerating dissolution. 
For the 7\% decane group, though the PDF in Figure \ref{fgr:SC2}A was similar to other groups, we observed a slightly slower reaction when compared to the lower concentrations.
It is possibly due to the slightly larger droplet size of 7\% decane group (5-7$\mu$m) that contributed to lower dissolution rate.

\subsection{Theoretical analysis and discussion}
The scaling of the dissolution of reacting droplets with $R^{-n}$ ($n\sim2$) is rationalized as below.
The biphasic reaction between alkali and OA mainly occurs at the surface of droplets, as the solubility of OA in water is low, and the solubility of alkali in OA droplets is also low. 
Considering a droplet on the substrate in the shape of a spherical cap, the droplet shrinkage is assumed to be in a constant contact angle (CA) mode for simplicity, i.e., the droplet keeps its contact angle constant all the time \cite{zhang2015mixed}.
Diffusion-dominated dissolution of the droplet is schematically demonstrated in Figure S3.

Based on the established transport-reaction model in our previous work \cite{li2020speeding}, the transport of the product from the droplet surface to bulk obeys Prandtl-Blasius-Pohlhausen-type behaviour in a laminar flow.
At given time $t$, the thickness of the boundary layer of the product concentration around the droplet is inversely proportional to the droplet size $R_{(t)}$ \cite{zhang2015formation}.

\begin{equation}
\label{eq:diffusion1}
\rho R_{(t)} \dot R\sim D\sqrt{Pe}c_{pr,sur}.
\end{equation}

\begin{equation}
\label{eq:diffusion2}
\dot R \propto R_{(t)}^{-1} c_{pr,sur}.
\end{equation}

Here $c_{pr,sur}$ is the concentration of the product at the droplet surface. 
The product density $\rho$ and the diffusion coefficient $D$ are constants.
$R_{(t)}$, and $\dot R$ are respectively  the lateral radius of the droplet at $t$ and the rate of change in the radius with time. Peclet number ($Pe$) of the bulk flow is defined as $Pe=Q/(wD)$, where $Q$ and $w$ are the volume flow rate and the channel width.

At given time $t$, the production rate of product $\dot m_{pr}$ can be expressed in terms of the interfacial area $A_{(t)}$ \cite{fallah2014enhanced,dyett2020accelerated,mondal2018enhancement}.

\begin{equation}
\label{eq:production1}
  \dot m_{pr} \propto A_{(t)} \cdot k \cdot [OA] [OH^-]_{sur}.
\end{equation}
$k$ is the reaction rate coefficient.

As OA is insoluble in water, the concentration of OA at the droplet surface is assumed to be constant with time.
The concentration of alkali at the droplet surface $[OH]^-_{sur}$ is determined by the $Pe$-dependent transport and reaction rate coefficient $k$ \cite{li2020speeding}.
The level of $[OA]$, $[OH]^-_{sur}$ or $k$ does not depend on the droplet size. 
The surface area of the droplet $A_{(t)}$ is $\sim R_{(t)}^2$, hence
\begin{equation}
\label{eq:production2}
 \dot m_{pr} \propto R_{(t)}^{2} \cdot k \cdot [OA] [OH^-]_{sur}.
\end{equation}

The concentration of the product at the droplet surface is mainly balanced by three factors: reaction rate coefficient $k$, product diffused into the droplet, and by the flow from the droplet.
The product diffuses into the droplet and builds up the concentration inside the droplet $c_{pr,in}$.
The liquid volume $V_{(t)}$ in the droplet is $\sim R_{(t)}^3$, $V_{(t)} \propto R_{(t)}^3$.
Coupling with equation \ref{eq:production2}, the concentration of the product inside the droplet was given as
\begin{equation}
\label{eq:inC}
 c_{pr,in} \propto \dot m_{pr}/V_{(t)} \propto R_{(t)}^{-2} \cdot k \cdot [OA] [OH^-]_{sur} / R_{(t)}^{-3} = R_{(t)}^{-1} \cdot k \cdot [OA] [OH^-]_{sur}.
\end{equation}

As the droplet size is small with tens to hundreds nanometer in height and the product is soluble in droplet liquid, the concentration of the product at the droplet surface $c_{pr,sur}$ is same as that inside the droplet $c_{pr,in}$.

\begin{equation}
\label{eq:surfaceC}
 c_{pr,in} \sim c_{pr,sur}.
\end{equation}

From equations \ref{eq:diffusion2}, \ref{eq:inC}, and \ref{eq:surfaceC} we obtain the scaling law for the droplet dissolution rate $\dot R$ as function of droplet size $R_{(t)}$ as

\begin{equation}
\label{eq:scaling}
 \dot R \propto R_{(t)}^{-1}c_{pr,sur} \propto R_{(t)}^{-1} \cdot R_{(t)}^{-1} \cdot k \cdot [OA] [OH^-]_{sur} = R_{(t)}^{-2} \cdot k \cdot [OA] [OH^-]_{sur}.
\end{equation}

Equation \ref{eq:scaling} is in agreement with our experimental results, that the dissolution rate $\dot R\propto R^{-2}$.

Our recent work reported the influence of external flow Peclet number on dissolution of reaction droplets \cite{li2020speeding}.
The rate of droplet dissolution scales with $Pe^{3/2}$.
Therein the dependence of the reaction rate on the droplet size could be neglected in both experimental and theoretical analysis, as the dissolution rate was represented by the surface coverage, an average over many droplets. 
In this work, reaction rate is analyzed on an individual droplet level.
Together, these two studies provide a comprehensive understanding on the dependence of droplet reaction on droplet size and on the reactant flow.

Enhanced chemical kinetics in small droplets has been frequently reported in literature.\cite{nam2017abiotic,fallah2014enhanced,ingram2016going}
However, the exact mechanism for the enhancement depends on the type of reactions and the configuration of the droplets. 
For flying droplets from atomized spray, all reactants were compartmentalized inside the droplet, in contrast to our current systems where reactants are in two immiscible phases. 
Enhanced electron transfer and molecular configuration at the droplet interface led to a low active energy barrier and thus a large rate coefficient k, attributed to the accelerated reaction rate of droplets on-flight.\cite{lee2015microdroplet,nam2017abiotic} However, the enhanced kinetics in our system was due to the high surface-to-volume ratio of small droplets and size-dependent mass transport of the product, while the intrinsic kinetics of chemical reaction itself ($k$) remains constant.

For gas-evolution reactions between droplets and a reactant dissolved in a stationary surrounding phase, the gas production rate was reproduced to be also higher in smaller droplets. However, the gas concentration inside droplets is inversely proportional to droplet radius \cite{dyett2020accelerated,dyett2019plasmonic}, different from $R^n (n\sim2)$ for droplet dissolution in this work. The reason for a stronger dependence on the droplet size in our reaction is attributed to the influence of the droplet size on the transport of the product to the external flow in equation \ref{eq:diffusion2}, and to the effect of surface-to-volume ratio on the product concentration in droplets as shown in equation \ref{eq:inC}.
The shifted balance between chemical reaction and enhanced mass transport contributed to faster overall kinetics of the process.

The results and analysis based on our model reaction are applicable to other reactions with reactants separated in droplets and in the flow of the immiscible phase where the product transport is the rate-limiting step.

\section{Conclusions}
This work reports the quantitative correlation between the chemical reaction rates and the size of reacting nanodroplets.
The model reaction in our study is neutralization between oleic acid droplets on a solid surface and alkali in an external flow.
The dissolution of the product results in the shrinkage of the droplet size, which reveals the reaction rate at the droplet surface in time.
The quantitative analysis reveals that the reaction rate increases with the decrease in the droplet size, regardless of the coverage of the droplets on the surface, the flow rate of the reactant solution, or the addition of non-reactive liquid in the droplets.
The overall kinetics scales with $\dot R\propto R^{-2}$.
The scaling law is attributed to faster accumulation of the product in the droplets from the larger surface to volume ratio at smaller droplet size and effect of droplet size on the product transport in the flow.

The finding in this work has advanced the current understanding of reaction kinetics of droplets, complementary to the studies on reactions in flying droplets \cite{bain2016accelerated,lai2018microdroplets,nam2018abiotic,lee2019micrometer,marsh2019reaction,wei2020accelerated}, and in immersed reacting droplets without an external flow \cite{fallah2014enhanced,dyett2020accelerated,narayan2005water}.
The insights from this work may be valuable for designing and controlling biphasic reactions in droplets systems with a reactant in the external flow.

\begin{acknowledgement}
The authors acknowledge the support from the Natural Science and Engineering Research Council of Canada (NSERC), Future Energy Systems (Canada First Research Excellence Fund), and the start-up fund from Faculty of Engineering, University of Alberta.
The authors appreciate the funding from the Canada Research Chairs program.
A. K. is JSPS Overseas Research Fellow.
\end{acknowledgement}

\begin{suppinfo}
The following files are available free of charge.
\begin{itemize}
  \item Supporting Information: Figure S1; Figure S2; Figure S3; Chemicals data sheet.
\end{itemize}
\end{suppinfo}

\bibliography{Citation}

\end{document}